\documentclass[prb,aps,amsmath,preprint]{revtex4}
\usepackage{graphicx}
\usepackage[version=3]{mhchem}
\usepackage{dcolumn}
\usepackage{multirow}
\usepackage{bm}
\usepackage{subfigure}
\usepackage{amssymb}
\usepackage{color}

\begin{document}
\title{High pressure structural, elastic and vibrational properties of green energetic oxidizer ammonium dinitramide}
\author{N. Yedukondalu$^1$, Vikas D. Ghule$^2$ and G. Vaitheeswaran$^{1*}$ }
\affiliation{$^1$Advanced Centre of Research in High Energy Materials (ACRHEM),
University of Hyderabad, Prof. C. R. Rao Road, Gachibowli, Telangana, Hyderabad-500046, India.\\
$^2$Department of Chemistry, National Institute of Technology, Kurukshetra 136119, Haryana, India.
}
\date{\today}
\begin{abstract}
Ammonium DiNitramide (ADN) is one of the most promising green energetic oxidizer for future rocket propellant formulations. In the present work, we report a detailed theoretical study on structural, elastic and vibrational properties of the emerging oxidizer under hydrostatic compression using various dispersion correction methods to capture weak intermolecular (van der Waals $\&$ hydrogen bonding) interactions. The calculated ground state lattice parameters, axial compressibilities and equation of state are in good accord with the available experimental results. Strength of intermolecular interactions has been correlated using the calculated compressibility curves and elastic moduli. Apart from this, we also observe discontinuities in the structural parameters and elastic constants as a function of pressure. Pictorial representation and quantification of intermolecular interactions are described by the 3D Hirshfeld surfaces and 2D finger print maps. In addition, the computed infra-red (IR) spectra at ambient pressure reveal that ADN is found to have more hygroscopic nature over Ammonium Perchlorate (AP) due to the presence of strong hydrogen bonding. Pressure dependent IR spectra show blue- and red-shift of bending and stretching frequencies which leads to weakening and strengthening of the hydrogen bonding below and above 5 GPa, respectively. The abrupt changes in the calculated structural, mechanical and IR spectra suggest that ADN might undergo a first order structural transformation to a high pressure phase around 5-6 GPa. From the predicted detonation properties, ADN is found to have high and low performance characteristics (D$_{CJ}$ = 8.09 km/s and P$_{CJ}$ = 25.54 GPa) when compared with ammonium based energetic oxidizers (D$_{CJ}$ = 6.50 km/s and P$_{CJ}$ = 17.64 GPa for AP, D$_{CJ}$ = 7.28 km/s and P$_{CJ}$ = 18.71 GPa for ammonium nitrate) and well-known secondary explosives for which D$_{CJ}$ = $\sim$ 8-10 km/s and P$_{CJ}$ = $\sim$ 30-50 GPa, respectively.
\end{abstract}

\maketitle
\section {Introduction}
Investigation and development of low toxicity and eco-friendly green energetic (primary, secondary explosives, oxidizers, and propellants) materials become an emerging field of research during the last decade. Especially, usage of green energetic oxidizers and propellants for pyrotechnic formulations prevents environmental pollution.\cite{Steinhauser} Ammonium Perchlorate (AP, NH$_4$ClO$_4$) and Ammonium Nitrate (AN, NH$_4$NO$_3$) are widely used energetic oxidizers. However, AP produces HCl as a combustion product; for instance, 503 tonnes of propellant containing AP liberates 100 tonnes of HCl and other chlorine containing compounds during its burning, thereby causing ``Ozone layer depletion" in the stratosphere.\cite{Venkatachalam1} Also, the large amount of HCl could cause ``acid rain" and high concentrations of perchlorate can affect the function of the thyroid gland.\cite{Venkatachalam1} Consequently, AN was considered to be an environment-friendly alternative to AP but its extreme low enthalpy of formation and relatively low density prevents AN usage in solid composite propellants (SCP).\cite{Manelis} Therefore, there is a significant interest in finding environmentally benign replacements for the extensively used AP and AN. Ammonium DiNitramide (ADN, NH$_4$N(NO$_2$)$_2$) has been identified as a promising new green energetic oxidizer for solid rocket propellants because of the desirable properties namely low sensitivity of ammonium salts, high-performance characteristics of nitramine compounds, high oxygen balance and absence of chlorine atoms.\cite{Cui} Apart from this, a shuttle can transport 8$\%$ more mass into orbit by using ADN as a propellant in place of AP.\cite{Badgujar}
\par On the other hand, high pressure/temperature polymorphs and structural phase transitions have much impact on the fundamental properties such as intermolecular interactions, chemical bonding, crystal structures, thermo-elastic properties of energetic molecular solids.\cite{Davidson1} Two different energetic polymorphs display distinct energetic performance attributing different crystal density thereby changing in the detonation characteristics. Therefore, there is a significant interest in understanding the phase stability and polymorphism of energetic materials under extreme conditions. Extensive reports are available in the literature addressing the high temperature behavior and thermal decomposition of ADN.\cite{Tompa,Yang,Oxley,Libbecke,Matsunaga1,Matsunaga2} Russell et al\cite{Russell} reported the Pressure-Temperature (P-T) diagram of ADN up to 10 GPa and 198-398 K and a reversible phase transition is observed from $\alpha$ $\rightarrow$ $\beta$ at around 2.0 GPa. In contrast to the previous study,\cite{Russell} $\alpha$-ADN is found to be stable up to 4.03 GPa.\cite{Davidson3} To complement the experiments, atomistic simulations are an effective way to model the crystal structures and their physical $\&$ chemical properties. Zhu et al\cite{Zhu1} made a comparative study of electronic, vibrational and thermodynamic properties between AP and ADN at ambient pressure. Apart from this, Sorescu et al\cite{Sorescu1,Sorescu2} investigated the structural and electronic properties of ADN using plane wave pseudo potential approach at ambient as well as at high pressure (0-600 GPa) without treating intermolecular interactions. They found that ambient phase of ADN transforms to triclinic (P$\bar{1}$) structure around 10 GPa.\cite{Sorescu2} However, the examined compound is an ionic-molecular solid and hence contribution towards intermolecular interaction from dispersion forces may be quite low but the ions are linked through hydrogen bonding networks. Therefore, investigation of this material by treating intermolecular interactions at ambient as well as at moderate pressures is crucial to obtain a fundamental knowledge at the atomistic level. With this motivation, in the present work, we have studied the pressure dependent structural, elastic and vibrational properties of ADN using dispersion corrected DFT-D2 method.
\section{Computational details}
First principles calculations were accomplished with the projector augmented wave (PAW) method as implemented in VASP package.\cite{Kresse} Generalized gradient approximation (GGA) in the Perdew-Burke-Ernzerhof (PBE) parametrization was considered as the exchange-correlation functional to treat electron-electron interactions.\cite{burke} Structural and elastic properties were calculated by setting the convergence criteria below 1.0 $\times$ 10$^{-8}$ eV for total energies, residual forces to be less than 1.0 $\times$10$^{-4}$ eV/$\AA$ and stresses are limited to 0.02 GPa. To treat the missing dispersion interactions, we have used various recently developed methods namely additive pair-wise D2,\cite{Grimme} TS,\cite{Tkatchenko1} TS-SCS, \cite{Tkatchenko2} and non-local (vdW-DF)\cite{Dion} correction methods as implemented in the VASP code. The density functional perturbation theory (DFPT) calculations were performed using plane wave  pseudo potential (PW-PP) approach incorporated through CASTEP package.\cite{Payne} Norm conserving (NC)\cite {Troullier} PW-PPs were used to calculate the lattice dynamical properties. A kinetic energy cutoff of 950 eV and 2$\pi\times$0.04 $\AA^{-1}$ separation of k-mesh according to the Monkhorst-Pack grid scheme were used in the calculations.\cite{Monkhorst} The self-consistent energy convergence and maximum force between atoms were set to 5.0$\times$10$^{-6}$ eV/atom and 0.01 eV/$\AA$, respectively. The maximum displacement and stress were set to be 5.0$\times$10$^{-4}\AA$ and 0.02 GPa, respectively. Hirshfeld surfaces\cite{McKinnon} and 2D finger print maps\cite{Spackman} are calculated using CrystalExplorer 3.1.\cite{www} 

\section{Results and discussion}
\subsection{Crystal structure and equation of state}
Gilardi et al\cite{Gilardi} reported that ADN crystallizes in the primitive monoclinic structure having space group $P2_1/c$ with 4 molecules per unit cell at ambient conditions. The crystal structure and atomic labels corresponding to each inequivalent atom for one molecule present in the unit cell of ADN are given in figure \ref{struct}. In contrast to the two dimensional network of AN,\cite{} the hydrogen bonds in ADN are directed tetrahedrally to form a three dimensional hydrogen bonded networks. ADN contains two independent three dimensional networks of hydrogen bonds.\cite{Gilardi} The first one involves a hydrogen bonding chain propagating along c-axis by connecting one of the nitro group (N3-O3A-O3B) of the anion via O3A to hydrogen atoms H2 and H3 of the NH$_4$ cation (see figure \ref{ADN-H}a). The second one associated with the anion consists of twisted O2A which is out of plane for the first nitro group of the anion to form hydrogen bonding through H1, H4 atoms and the adjacent layers of cation form a helical structure along ‘b’-axis (see figure \ref{ADN-H}b). The two inter-penetrating patterns form two independent three dimensional hydrogen bonding networks as presented in figure \ref{ADN-H}. In order to get the equilibrium ground state structure of ADN, we first performed the full structural optimization using PBE-GGA functional, various dispersion DFT-D (DFT-D2, vdW-TS, TS-SCS) and non-local vdW-DF correction methods by starting with single crystal X-ray diffraction data.\cite{Gilardi} The obtained equilibrium volume with PBE-GGA functional is overestimated by around 13.2 $\%$. While the predicted volumes are found to differ by +0.001 $\%$, +4.0 $\%$, +5.1 $\%$ and +0.003 $\%$ using DFT-D2, vdW-TS, TS-SCS and vdW-DF methods, respectively. Here '+' represents an overestimation of equilibrium volume when compared to the experiments.\cite{Gilardi} Overall, we find a good agreement between our calculated equilibrium volume using D2 and vdW-DF methods when compared with the experiments and all the results are presented in Table \ref{tab:table1}.

\par Further, the obtained ground state structures at ambient pressure using DFT-D2 method were used to perform high pressure calculations in the pressure range 0-5 GPa in steps of 0.5 GPa. The computed lattice constants are found to decrease monotonically with pressure up to 5 GPa while the monoclinic angle ($\beta$) which is increasing with pressure as depicted in figure \ref{abc-ADN} in the pressure range 0-5 GPa. Russell et al\cite{Russell} reported the P-T phase diagram of ADN using energy dispersive X-ray diffraction and Raman spectroscopic measurements under pressure up to 10 GPa and they claimed that $\alpha$-ADN (ambient phase of ADN) transforms to $\beta$-ADN phase through a first-order polymorphic phase transition around 2 GPa. However, Davidson and co-workers \cite{Davidson3} revisited the high pressure behavior of ADN using a combination of X-ray and Neutron powder diffraction techniques and found that ADN is quite stable up to 4.03 GPa. From the theoretical perspective, without treating intermolecular interactions, Sorescu et al\cite{Sorescu2} predicted that ADN is stable up to 10 GPa and it transforms from monoclinic ($P2_1/c$) to triclinic ($P\bar{1}$) symmetry above 10 GPa. Thus there exists an inconsistency between various studies regarding the high pressure behavior of ADN. In order to resolve this issue, we have extended our pressure range from 5-10 GPa in our calculations using DFT-D2 method. As illustrated in the inset of figures \ref{abc-ADN}a-e, we could see a discontinuity in the lattice constants {\bf ‘a’}, {\bf ‘c’} and monoclinic angle {$\bf \beta$} whereas lattice constant {\bf ‘b’} exhibits monotonic behavior as a function of pressure up to 10 GPa. The discontinuities in the structural properties of ADN might suggest a structural transition/distortion around 6 GPa. Similar kind of discontinuities were seen in the case of solid nitromethane for the lattice constants and bond parameters using D2 method which discloses the structural phase transition in the pressure range 10-12 GPa.\cite{kondaiah}
\par As illustrated in figure \ref{abc-ADN}e, the volume is decreasing with pressure and a discontinuity is observed in the pressure range 6-7 GPa. Moreover, we have also computed the equilibrium bulk modulus (B$_0$) and its pressure derivative (B$_0'$) by fitting the pressure-volume data (0-5 GPa) to 3$^{rd}$ order Birch-Murnaghan equation of state.\cite{Birch} The obtained B$_0$ and B$_0'$ values using PBE, various DFT-D and vdw-DF methods are presented in Table \ref{tab:table1} along with the experimental results.\cite{Davidson3,Peiris} It is found that the obtained B$_0$ and B$_0'$ values are consistent with the experimental data \cite{Davidson3,Peiris} and other theoretical calculations.\cite{Sorescu2} The calculated B$_0$ values for ADN using DFT-D and vdw-DF methods are larger than nitromethane (8.3 GPa)\cite{Cromer,Citroni}, $\alpha$-RDX (13.9 GPa)\cite{Yoo1}, $\beta$-HMX (15.7 GPa)\cite{Yoo2} and PETN-I (12.3 GPa)\cite{Yoo1} which indicates that ADN is harder than the conventional secondary explosives. Apart from this, we have also calculated the normalized lattice constants and the data are presented in figure \ref{abc-ADN}f. The lattice constants \textbf{a}, \textbf{b} and \textbf{c} shrink with different compressibilities 96.5, 92.70 and 96.49 $\%$, respectively for ADN in the studied pressure range of 0-5 GPa. The calculated axial compressibilities are found to be consistent with the X-ray powder diffraction data.\cite{Davidson3} Also, the axial compressibilities show that {$\bf b$}-axis is the most compressible over {$\bf a$}- and {$\bf c$}-axes for ADN, which is due to weaker intermolecular interactions (van der Waals and/or hydrogen bonding) along the {$\bf b$}-axis as shown in figure \ref{ADN-H}b.

\subsection{Intermolecular interactions-Hirshfeld surface and 2D finger print maps}
X-ray diffraction\cite{Gilardi} and spectroscopic\cite{Cui} studies reveal that hydrogen bonding plays a vital role in determining the stability of ADN. In order to understand the effect of pressure on hydrogen bonding, we have plotted the normalized intermolecular hydrogen bond parameters (bond lengths and angles) as a function of pressure as depicted in figure \ref{BLA-H}. H3...O3A and H4...O2B intermolecular bonds are more compressible over H1...O2A and H2...O3A bonds. ADN lattice is found to be more compressible along crystallographic {\bf b}-direction which is due to large compressibility (mainly arises from H4...O2B) of helically structured hydrogen bonding network along the {\bf b}-axis (see figure \ref{ADN-H}b). We also observe sharp discontinuities in the hydrogen bond parameters in the pressure range between 6-7 GPa. The sharp increase in hydrogen bond lengths may lead to a weakening of the hydrogen bonding which can be further analyzed through the calculated Hirshfeld surface, 2D finger print maps and IR spectra in Secs. III B and III D, respectively. In addition, we have also calculated the intra-molecular interactions of both cation (NH$_4^+$) and anion (N(NO$_2)_2^-$) as a function of pressure as illustrated in figure \ref{BLA}. Overall, the intermolecular interactions are significantly affected by pressure over intra-molecular interactions as depicted in figures \ref{BLA-H} and \ref{BLA}. The discontinuities in the bond parameters from both inter- and intra-molecular interactions may suggest a structural phase transition in ADN around 6 GPa.

\par Furthermore, pictorial representation and quantification of intermolecular interactions in molecular solids can also be described by 3D Hirshfeld surface (HS) in combination with 2D finger print maps. The HS and finger print maps are unique for a molecule in a crystal. Hirshfeld surfaces are produced through the partitioning of space within a crystal where the ratio of promolecule to procrystal electron densities is equal to 0.5, resulting in continuous non-overlapping surfaces. The molecular Hirshfeld surfaces of ADN are generated using standard (high) surface resolution with the 3D d$_{norm}$ surfaces were mapped over a fixed color range of -0.5 (red) to 0.5 $\AA$ (blue). The d$_{norm}$ surface is used for identification of very close intermolecular interactions. The d$_{norm}$ values are mapped onto the HS by using a red-blue-white color scheme: red, blue and white regions represent shorter, longer and the distance of contacts is exactly equals to the vdW radius separation, with positive, negative and zero d$_{norm}$ values, respectively. The calculated HS for various intermolecular interactions (O...H, N...H, O...O $\&$ N...O)  are presented in figure \ref{fig:HS} and the corresponding 2D finger print plots are illustrated in figure \ref{fig:2D}. In general, pair of "spikes" and "wing" features are identified as hydrogen bonding and van der Waals interactions, respectively. As illustrated in figures \ref{fig:2D}b, O...H/H...O contacts, which are attributed to N-O..H/N-H...O hydrogen bonding interactions appear as two sharp symmetric spikes in the 2D fingerprint maps. The O...H/H...O interactions provide the most significant contribution to the total Hirshfeld surface, accounting for 56.8 $\%$ (see figure \ref{fig:2D}b) of the total HS (see figure \ref{fig:2D}a). Similarly, the rest of intermolecular interactions N...H/H...N (10.0 $\%$), N...O/O...N (17.3 $\%$), O...O (11.9 $\%$), H...H (3.0 $\%$) and N...N (0.9 $\%$) with various proportions as shown in figure \ref{fig:2D}. In addition, we also attempted to investigate the contribution of various intermolecular interactions (see figure \ref{fig:2D}) to the HS as a function of pressure up to 10 GPa in steps of 1 GPa. As illustrated in figure \ref{fig:HS-P}, the contribution of H...O/O...H interactions decreases whereas N...H/H...N and O...O interactions increases to the total HS with progressing pressure (also see figure 2 of the supplementary material).\cite{support} In addition, we could also see an abrupt change in the contribution to the HS from various intermolecular interactions in the pressure range of 6-7 GPa (see figure \ref{fig:HS-P}) as inferred from the lattice and bond parameters (see figures \ref{abc-ADN}, \ref{BLA-H} $\&$ \ref{BLA}) of ADN.

\subsection{Elastic constants and mechanical stability}
Elasticity is a fundamental property of materials to examine their mechanical response to an applied stress. Quantifying and understanding the elastic properties of energetic materials is important to have a basic knowledge about the intermolecular interactions thereby analyzing stiffness of these materials. It has been previously reported \cite{Haycraft1} that stiffer the lattice less the sensitive it becomes to detonation from a mechanical shock initiation. Several investigations were carried out to measure the elastic stiffness constants for the well known secondary explosives namely RDX,\cite{Haycraft1} HMX,\cite{Stevens,Sun} PETN,\cite{Winey} and CL-20. \cite{Haycraft2} These studies reveal that overall RDX is the stiffest lattice which indicates the relative insensitiveness of RDX when compared to the above secondary explosives such as HMX, PETN and CL-20. On the similar path, to understand the mechanical response and stiffness of the examined solid energetic oxidizer, we have also calculated the elastic stiffness constants for ADN. It is well known from single crystal X-ray diffraction measurements,\cite{Gilardi} that ADN crystallizes in the monoclinic structure at ambient conditions and hence ADN possesses thirteen independent elastic constants. The obtained elastic constants at ambient pressure at D2 equilibrium volume are given in Table \ref{tab:table2}. Also, the calculated elastic constants meet the well known Born's stability criteria indicating that the investigated compound is mechanically stable at ambient pressure. In addition, the three diagonal elements can be used to correlate with the strength of intermolecular interactions along three crystallographic directions. Based on the measured stiffness constants, Haycraft \cite{Haycraft2} reported that CL-20 should be most sensitive to detonation along {$\bf a$}-axis and least sensitive along {$\bf b$}-axis. Also, it is observed that C$_{11}$ is the stiffest elastic constant for PETN which makes PETN to be least sensitive along [100] direction.\cite{Winey} Similarly, the calculated ordering of diagonal elastic constants for ADN as follows: C$_{11}$ $\textgreater$ C$_{33}$ $\textgreater$ C$_{22}$. C$_{11}$ is found to be the stiffest elastic constant for ADN, which is due to strong intermolecular interactions along {$\bf a$}-axis. This implies that ADN is less sensitive to detonation along [100] direction similar to that of PETN.\cite{Winey} The lower elastic moduli/high compressibilities reveal that ADN is found to be more sensitive to detonation along the {\bf b}-crystallographic direction.

\par Furthermore to understand the mechanical stability under hydrostatic pressure, we have also calculated the elastic constants as a function pressure and these are depicted in figure \ref{cij}a for the examined compound. As illustrated in figure \ref{cij}a, 11 out of 13 independent elastic constants are increasing with pressure while the remaining two C$_{66}$ and C$_{46}$ are softening under compression. In addition, we also observe discontinuities in the elastic stiffness constants for C$_{11}$ and C$_{33}$ as a function of pressure which are reflected from the pressure dependent lattice constants and bond parameters. Sinko and Smirnov\cite{Sinko1,Sinko2} proposed the theoretical conditions of elasticity under hydrostatic pressure. Liu et al\cite{Liu} derived the mechanical stability criteria for monoclinic crystal systems under pressure based on theoretical conditions of Sinko and Smirnov.\cite{Sinko1,Sinko2} We have also calculated the mechanical stability criteria as a function of pressure for the ambient phase of ADN. As illustrated in figure \ref{cij}b, the mechanical stability criteria (M6 = C$_{66}$-P ($>$ 0)) and (M9 = (C$_{33}$-P)(C$_{55}$-P)- C$_{35}^2$ ($>$ 0)) are violated above 6 GPa. This clearly indicates that ambient phase of ADN is found to mechanically unstable above 6 GPa. Therefore, these results indicate a possible structural phase transition in ADN around 6 GPa.

\subsection{Zone center phonons and IR spectra under high pressure}
In order to explore the dynamical stability, we have first calculated the zone center phonon frequencies for the investigated compound. ADN possesses monoclinic (P2$_1$/c) symmetry with Z = 4 f.u./cell, which results in 144 vibrational modes at the center of the Brillouin zone. The symmetry decomposition of the vibrational modes for the studied compound is as follows: $\Gamma_{tot}^{ADN}$ = 36A$_{u}$ $\oplus$ 36B$_{u}$ $\oplus$ 36A$_{g}$ $\oplus$ 36B$_{g}$. The detailed vibrational spectra analysis of each vibrational mode for ADN at ambient pressure can be found elsewhere. \cite{Zhu1}
\par To understand the hydrogen bonding and possible structural phase transformations in ADN, we have calculated the IR spectra at ambient as well as at high pressure. The calculated high frequency asymmetric and symmetric stretching bands of ADN in the near IR region are compared with AP at ambient pressure as presented in figure \ref{IR-0}. The stronger the hydrogen bonding is, the more is the displacement towards low frequency (red-shift) region that occurs. As illustrated in figure \ref{IR-0}, the high frequency N-H stretching modes of ADN show red-shift in comparison to the N-H stretching frequencies of AP which is consistent with the Fourier transform IR experiments.\cite{Cui} This clearly indicates that ADN has stronger hydrogen bonding than AP due to which ADN can bind large amounts of water by forming strong hydrogen bonding networks. This could be the reason why ADN is more hygroscopic than AP.\cite{Cui}

\par Further, IR spectra has been calculated up to 10 GPa in steps of 1 GPa. As illustrated in figure \ref{IR2}a, the far IR lattice mode frequencies which include both NH$_4$ and N(NO$_2)_2$ oscillations, NO$_2$ twisting, NH$_4$ translation, and rotation of NH$_4$ and N-NO$_2$ fragments of the ADN molecules are increasing with pressure below 5 GPa and these vibrational modes show red-shift above 5 GPa. The IR vibrational modes corresponding to torsional and asymmetric stretching of N-N and NO$_2$ bands exhibit blue-shift with pressure as depicted in figures \ref{IR2}b $\&$ c, which implies hardening of the lattice upon compression. Based up on blue and/or red-shift of IR frequencies and their corresponding intensities, Joseph et al\cite{Joseph} gave a consolidated description for strengthening and/or weakening of hydrogen bonding. Due to the presence of hydrogen bonding, it can be expected that the N-H strengthening frequency decreases (red-shift) with increasing pressure and this red-shift leads to a strengthening of hydrogen bonding.\cite{Reynolds} Also, our previous studies\cite{kondal,kondaiah1} and those of Pravica et al\cite{Pravica} on hydrogen bonded systems suggest that red-/blue-shift in the mid IR frequencies stabilize/destabilize the system under hydrostatic compression. As illustrated in figures \ref{IR2}d, e$ \&$ f, the N-H bending, symmetric and asymmetric stretching bands show blue-shift as a function of pressure up to 5 GPa which indicates the weakening of hydrogen bonding below 5 GPa. However, the N-H bending and symmetric bands show an abrupt red-shift and a sharp increase in intensity when compared to the IR spectra between the pressure regions 0-5 and 6-10 GPa for the investigated compound. The features of computed IR spectra under pressure is different below and above 5 GPa (see figures \ref{IR2}a, d, e$\&$ f). This is a clear spectroscopic indication of the weakening and strengthening of hydrogen bonding in ADN below and above 5 GPa, respectively. The weakening of hydrogen bonding below 5 GPa may suggest a structural transition in ADN. Strengthening of hydrogen bonding above 5 GPa reveals that the high-pressure phase has stronger hydrogen bonding nature when compared to the ambient phase. ADN is found to be stable up to 5 GPa which is in good accord with the recent X-ray and neutron diffraction study.\cite{Davidson3} While the abrupt changes in the structural, mechanical properties and weakening of hydrogen bonding strongly suggest that ADN undergoes a structural transformation around 6 GPa. There is an approximately 1 GPa deviation between structural, mechanical properties (discontinuities were observed between 6-7 GPa) and IR spectra (shift in the IR frequencies between 5-6 GPa) which is due to distinct pseudo potentials PAW and NC approaches were used in the present study.

\subsection{Detonation Properties}
The detonation characteristics namely detonation velocity and pressure are estimated using the EXPLO5 program which is based on the chemical equilibrium and steady-state model of detonation.\cite{Sućeska1} This program uses the Becker-Kistiakowsky-Wilson (BKW) semi-empirical equation of state to describe the state of gaseous detonation products and solid carbon is expressed by the Cowan and Fickett’s equation of state.\cite{Klapötke,Sućeska2} The obtained densities from ab-initio calculations with dispersion correction methods and the predicted solid state heats of formation (HOF) were used in the estimation of detonation properties. For AN, AP and ADN, the solid state HOFs were obtained by converting the calculated gas-phase HOFs of ions using the Born-Haber cycle and Jenkins approach.\cite{Jenkins} Computations for cations and anions were performed using the Gaussian09 program suite,\cite{Frisch} and the details are given in the supplementary material.\cite{support} The calculated crystal density, HOF, detonation velocity and pressure are presented in Table III along with the available experimental\cite{Yang} and theoretical results.\cite{Kraue,Nair,Zeng} The calculated detonation velocity and pressure at Chapman-Jouguet (CJ) point for the AN, AP and ADN oxidizers are relatively low (D$_{CJ}$ = $\sim$ 6-7 km/s and P$_{CJ}$ = $\sim$ 17-25 GPa) when compared to the high performance (D$_{CJ}$ = $\sim$ 8-10 km/s and P$_{CJ}$ = $\sim$ 30-50 GPa) conventional secondary explosives such as RDX, HMX, PETN, and ONC etc. The high detonation characteristics of ADN reveal its unique energetic nature among the three ammonium based oxidizers.


\section{CONCLUSIONS}
High pressure structural, mechanical and vibrational properties of ADN have been calculated using dispersion correction DFT-D and vdW-DF methods to capture weak inter-molecular interactions. The obtained ground state lattice parameters and equilibrium bulk moduli are in good agreement with the experimental results. The calculated compressibility curves and elastic moduli reveal that ADN is found to be more compressible along b-axis which is due to weak helical structured hydrogen bonding network along that axis. We also observe a discontinuity in the lattice constants, bond parameters, and elastic moduli as a function of pressure. The calculated IR spectra at ambient pressure reveal that ADN is found to be more hygroscopic nature than AP due to relatively strong hydrogen bonding nature. Pressure dependent IR spectra show blue- and red-shifts which lead to weakening and strengthening of hydrogen bonding below and above 5 GPa, respectively. Overall, the calculated structural, mechanical and IR spectra suggest that ADN undergoes a structural phase transition in the pressure range of 5-6 GPa. The crystal structure of the high pressure phase of ADN is an open challenge and it will be carried out in near future using crystal structure prediction method. The calculated detonation characteristics reveal that ADN is a powerful energetic oxidizer when compared to AN and AP. 

\section{Acknowledgments}
NYK would like to thank Defense Research and Development Organization (DRDO) through ACRHEM for the financial support under grant No. DRDO/02/0201/2011/00060:ACREHM-PHASE-II and the CMSD, University of Hyderabad, for providing computational facilities. \\
$^*$\emph{Author for Correspondence, E-mail: vaithee@uohyd.ac.in}

{}

{\pagestyle{empty}

\begin{figure*}[h]
\centering
\includegraphics[height = 4.0in,width=6.0in]{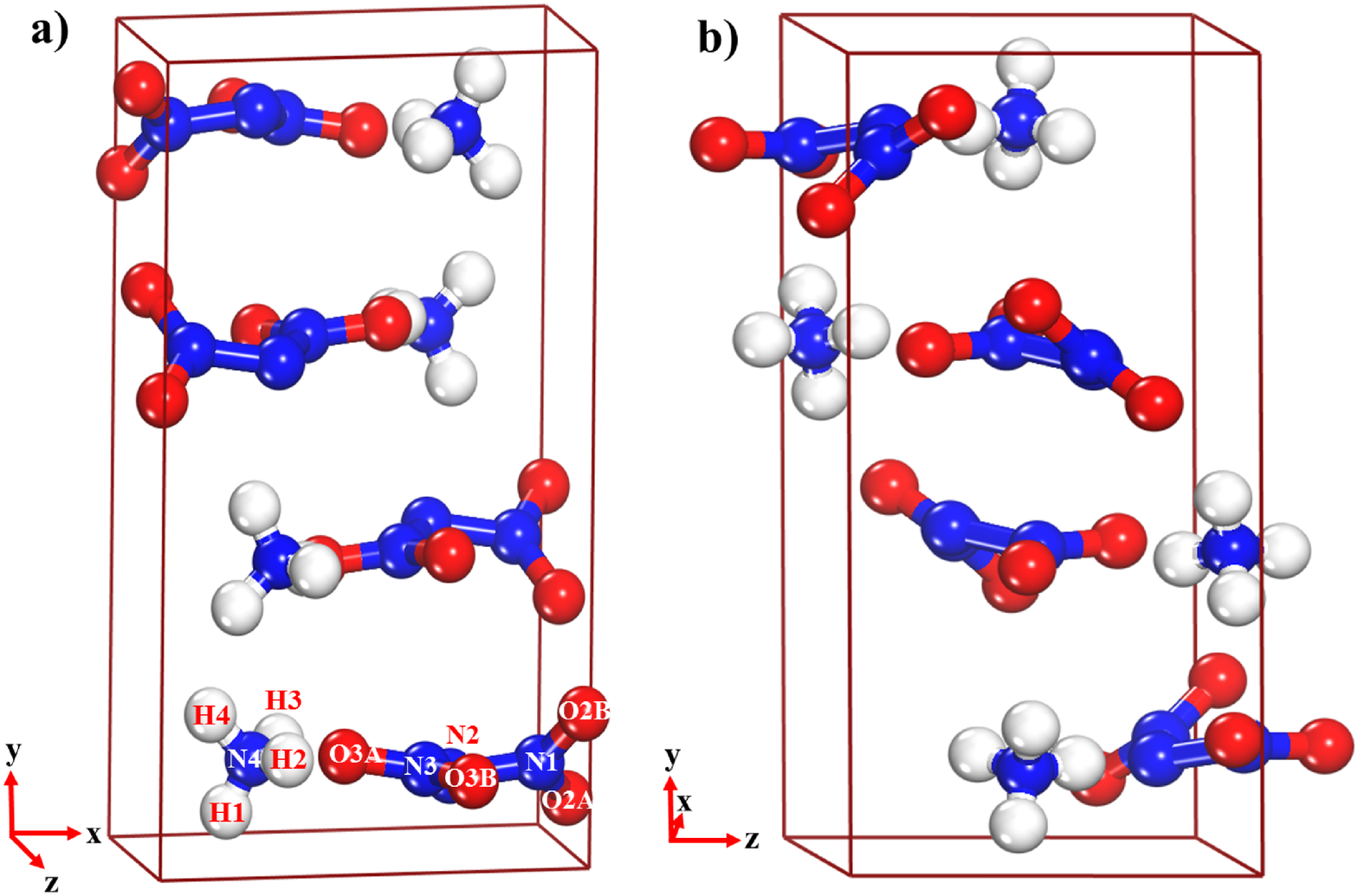}
\caption{(Color online) Crystal structure of ADN viewed along a) xy-plane and b) yz-plane. White, blue and red balls represent hydrogen, nitrogen and oxygen atoms, respectively.}
\label{struct}
\end{figure*}

\begin{figure*}[h]
\centering
\includegraphics[height = 7.8in,width=4.5in]{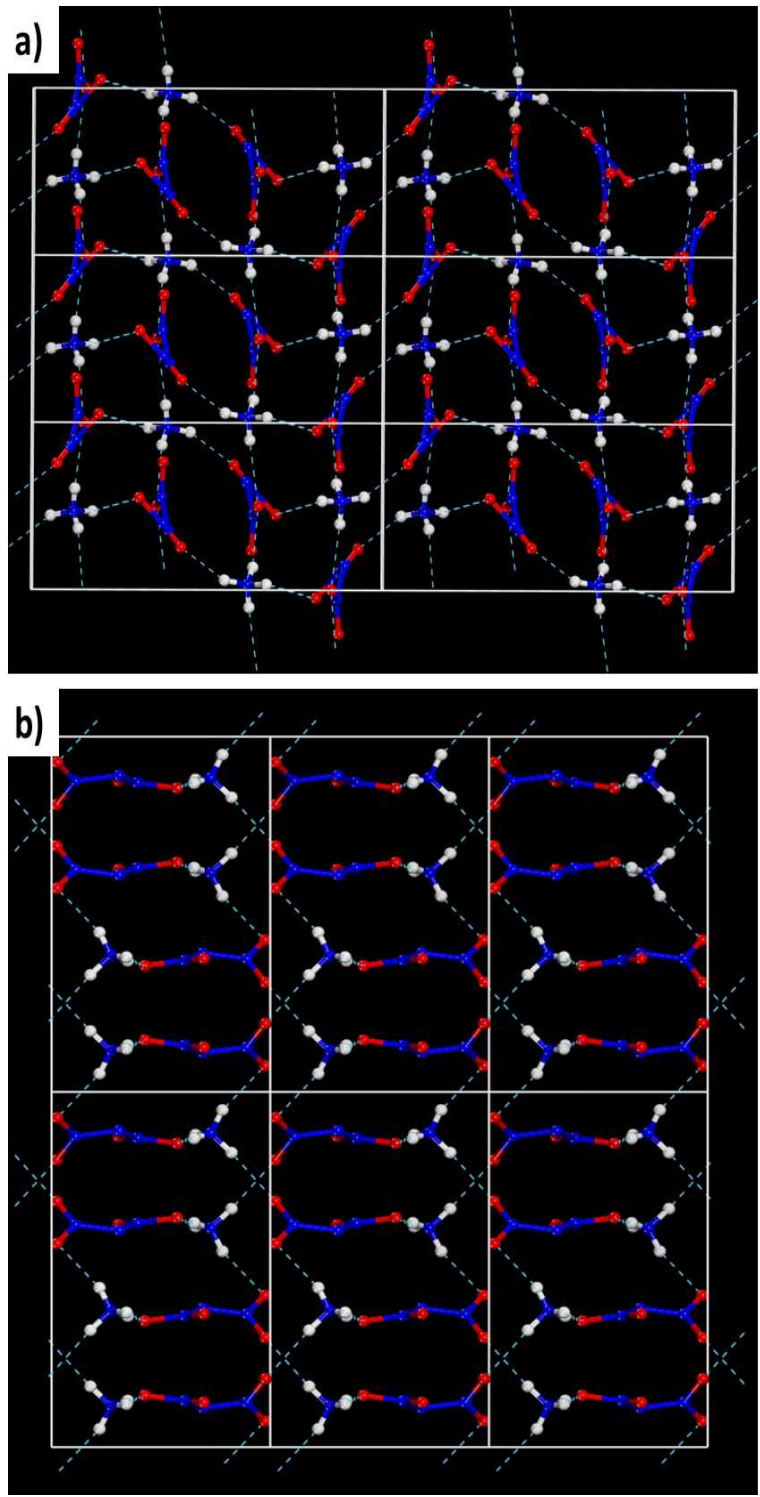}
\caption{(Color online) Three dimensional hydrogen bonding network in ADN as viewed along a) bc (yz) and b) ab (xy)-planes. White, blue and red color balls represent hydrogen, nitrogen and oxygen atoms respectively.}
\label{ADN-H}
\end{figure*}

\begin{figure*}[h]
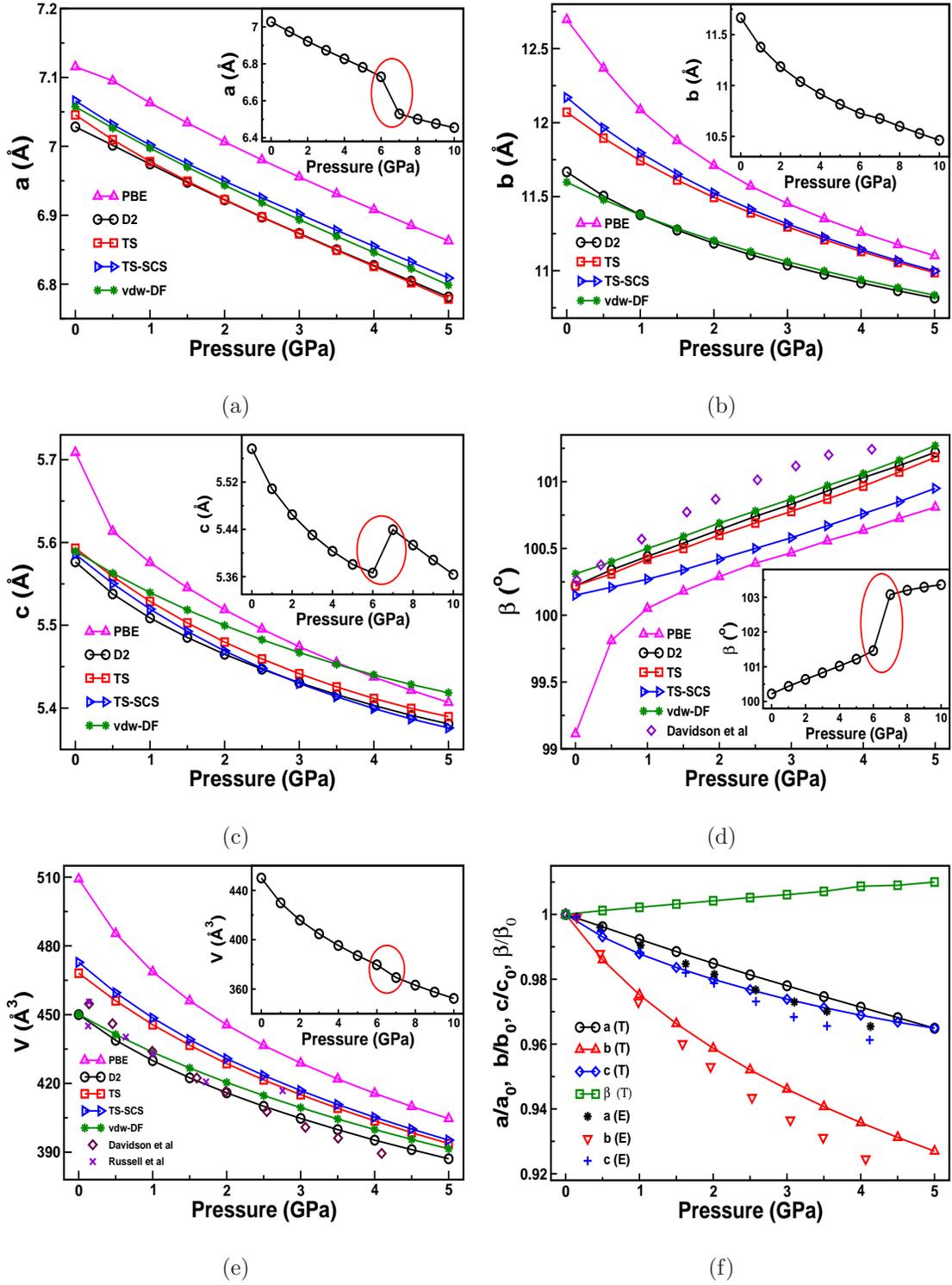

\centering
{\subfigure[]{\includegraphics[height = 2.2in,width=2.8in]{a_ADN.eps}}}  \hspace{0.05in}
{\subfigure[]{\includegraphics[height = 2.2in,width=2.8in]{b_ADN.eps}}}
{\subfigure[]{\includegraphics[height = 2.2in,width=2.8in]{c_ADN.eps}}}  \hspace{0.05in}
{\subfigure[]{\includegraphics[height = 2.2in,width=2.8in]{Beta_ADN.eps}}}
{\subfigure[]{\includegraphics[height = 2.2in,width=2.8in]{V_ADN.eps}}}  \hspace{0.05in}
{\subfigure[]{\includegraphics[height = 2.2in,width=2.8in]{NLP_ADN.eps}}}
\caption{(Color online) Calculated (a-e) lattice constants (a, b, c, $\beta$ and volume) and f) normalized lattice constants (DFT-D2) of ADN as a function of pressure up to 5 GPa using various DFT-D and non-local (vdW-DF) correction methods. Inset figures show the obtained lattice lattice constants under pressure up to 10 GPa using DFT-D2 method. Experimental data is taken from the Ref.\cite{Davidson3} Here T and E represent theory and experiment, respectively.}
\label{abc-ADN}
\end{figure*}

\begin{figure*}[h]
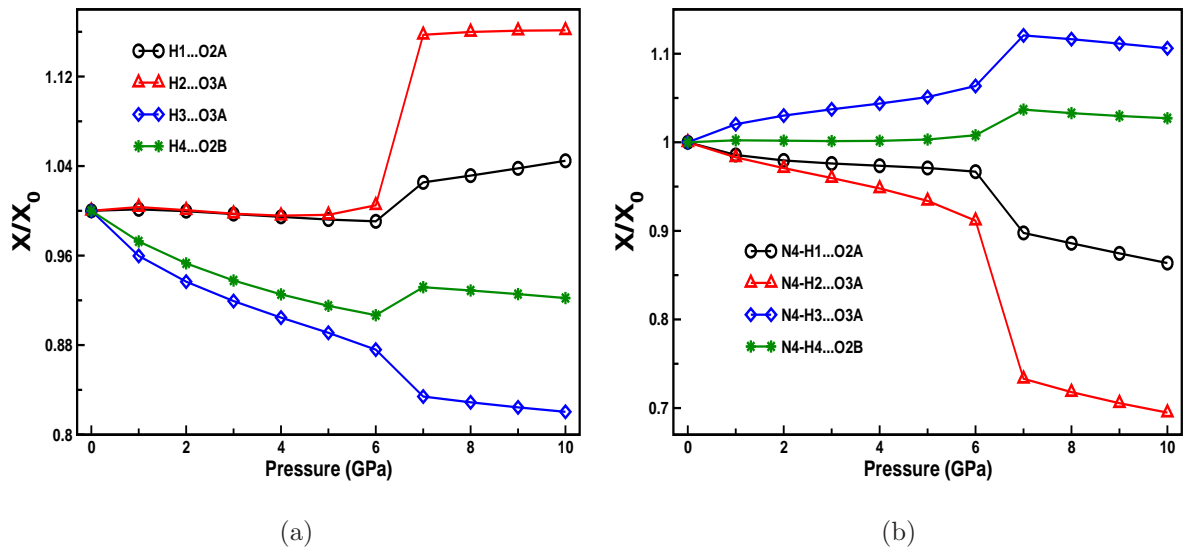

\centering
{\subfigure[]{\includegraphics[height = 2.5in,width=3.0in]{HBL.eps}}} \hspace{0.05in}
{\subfigure[]{\includegraphics[height = 2.5in,width=3.0in]{HBA.eps}}}
\caption{(Color online) Calculated normalized intermolecular hydrogen (a) bond lengths and (b) angles of ADN as a function of pressure. Where X$_0$ and X represent obtained bond parameters at ambient and as a function of pressure, respectively.}
\label{BLA-H}
\end{figure*}

\begin{figure*}[h]
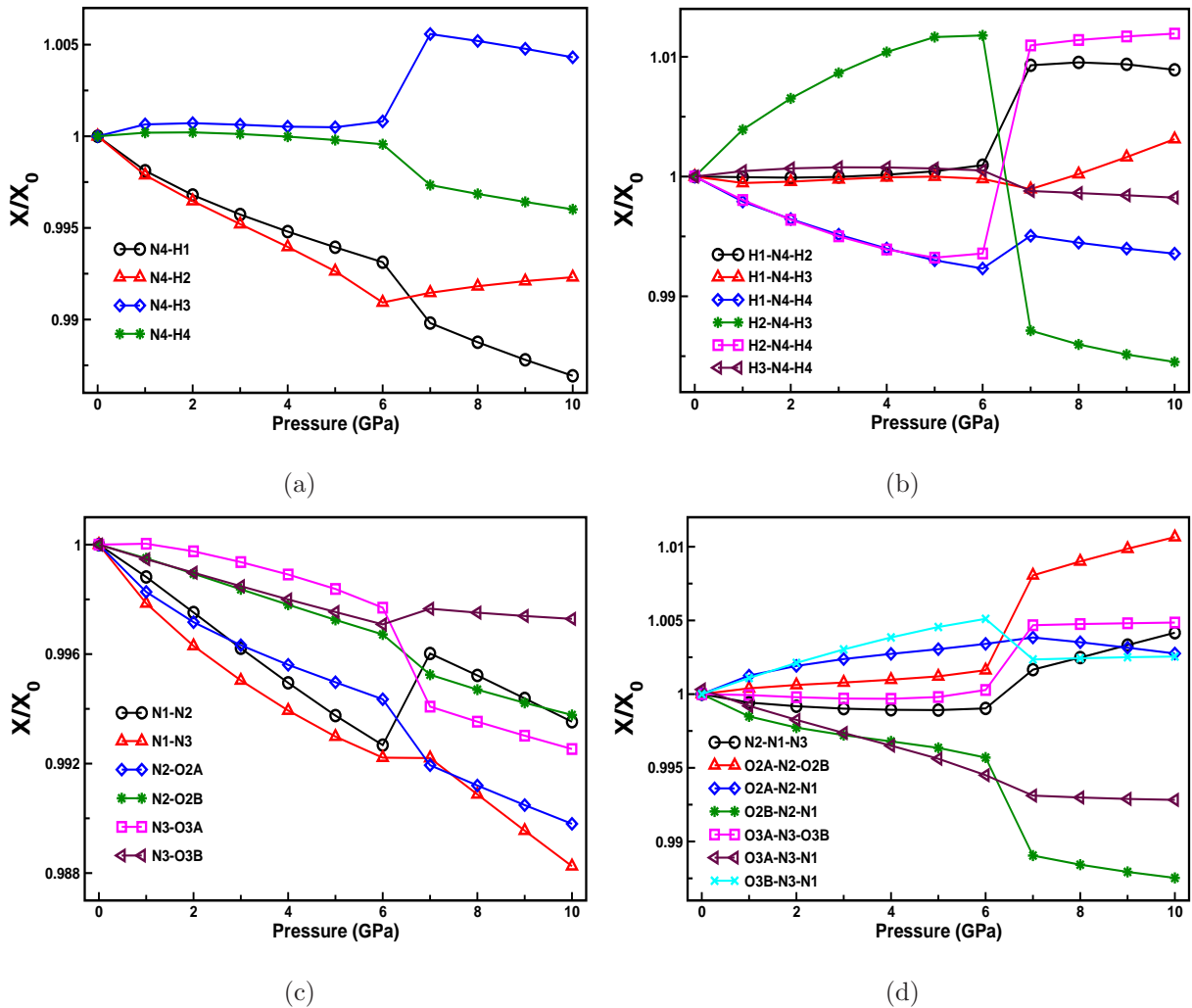

\centering
{\subfigure[]{\includegraphics[height = 2.3in,width=3.1in]{CBL.eps}}} \hspace{0.02in}
{\subfigure[]{\includegraphics[height = 2.3in,width=3.1in]{CBA.eps}}}
{\subfigure[]{\includegraphics[height = 2.3in,width=3.1in]{ABL.eps}}} \hspace{0.02in}
{\subfigure[]{\includegraphics[height = 2.3in,width=3.1in]{ABA.eps}}}
\caption{(Color online) Calculated normalized intra-molecular bond lengths (left) and angles (right) of (a,b) cation and (c,d) anion of ADN as a function of pressure. Where X$_0$ and X represent obtained bond parameters at ambient and as a function of pressure, respectively.}
\label{BLA}
\end{figure*}

\begin{figure*}[h]
\centering
\includegraphics[height = 8.0in,width=3.5in]{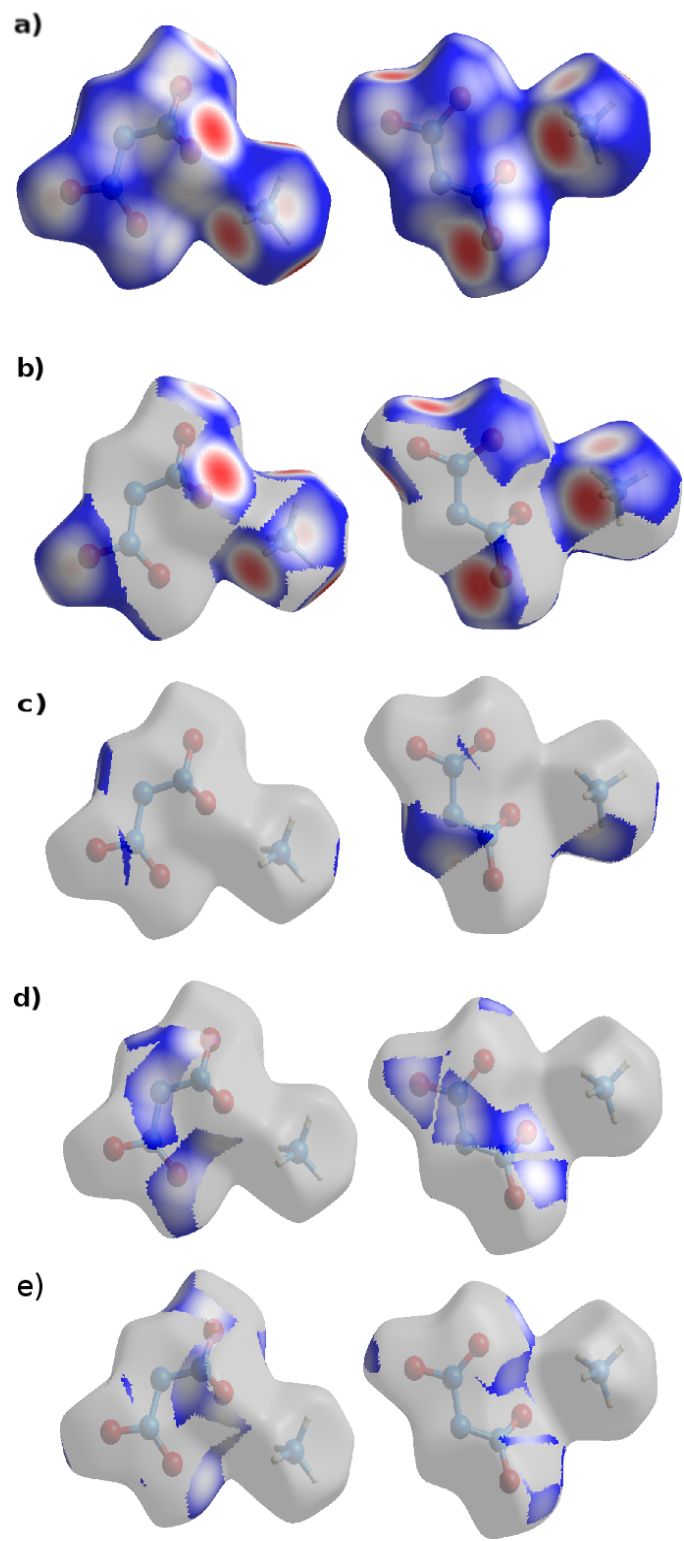}
\caption{(Color online) Hirshfeld surfaces of ADN displayed in front (left) and down (right) view obtained from a) total and various b) H...O/O...H, c) N...H/H...N, d) N...O/O...N $\&$ e) O...O intermolecular interactions.}
\label{fig:HS}
\end{figure*}

\begin{figure*}[h]
\centering
{\subfigure[]{\includegraphics[height = 2.5in,width=2.5in]{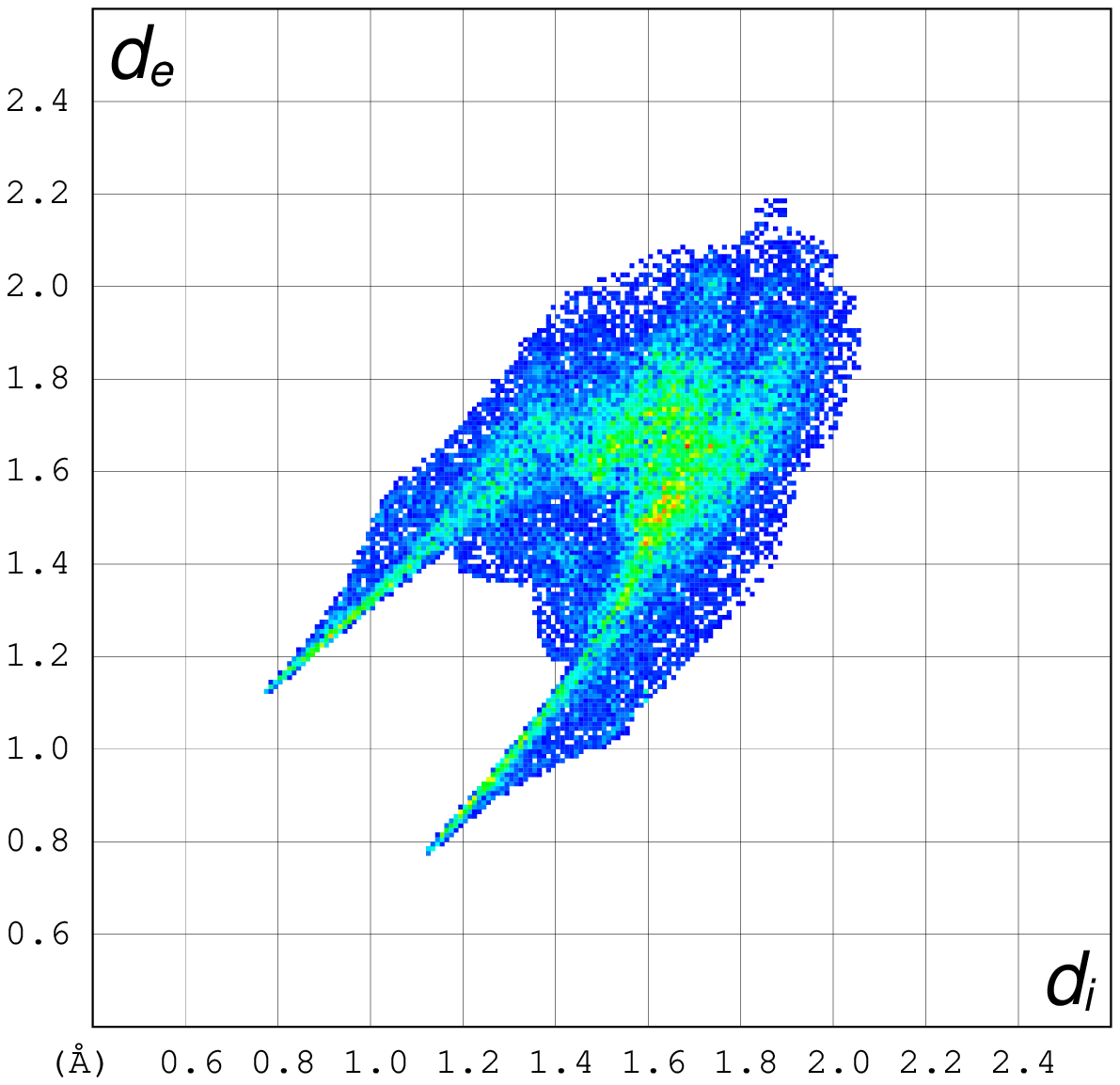}}} \hspace{0.1in}
{\subfigure[]{\includegraphics[height = 2.5in,width=2.5in]{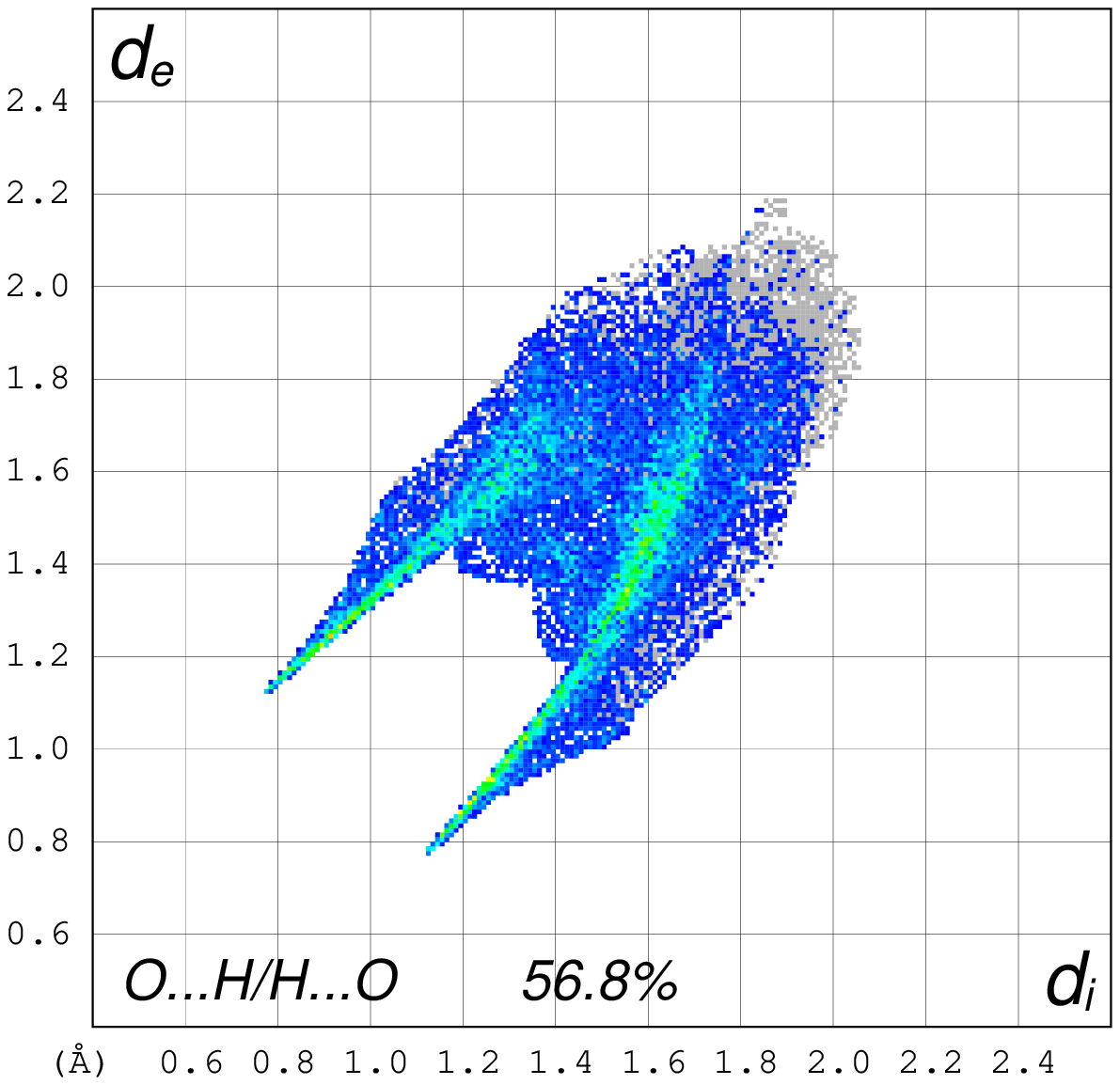}}}
{\subfigure[]{\includegraphics[height = 2.5in,width=2.5in]{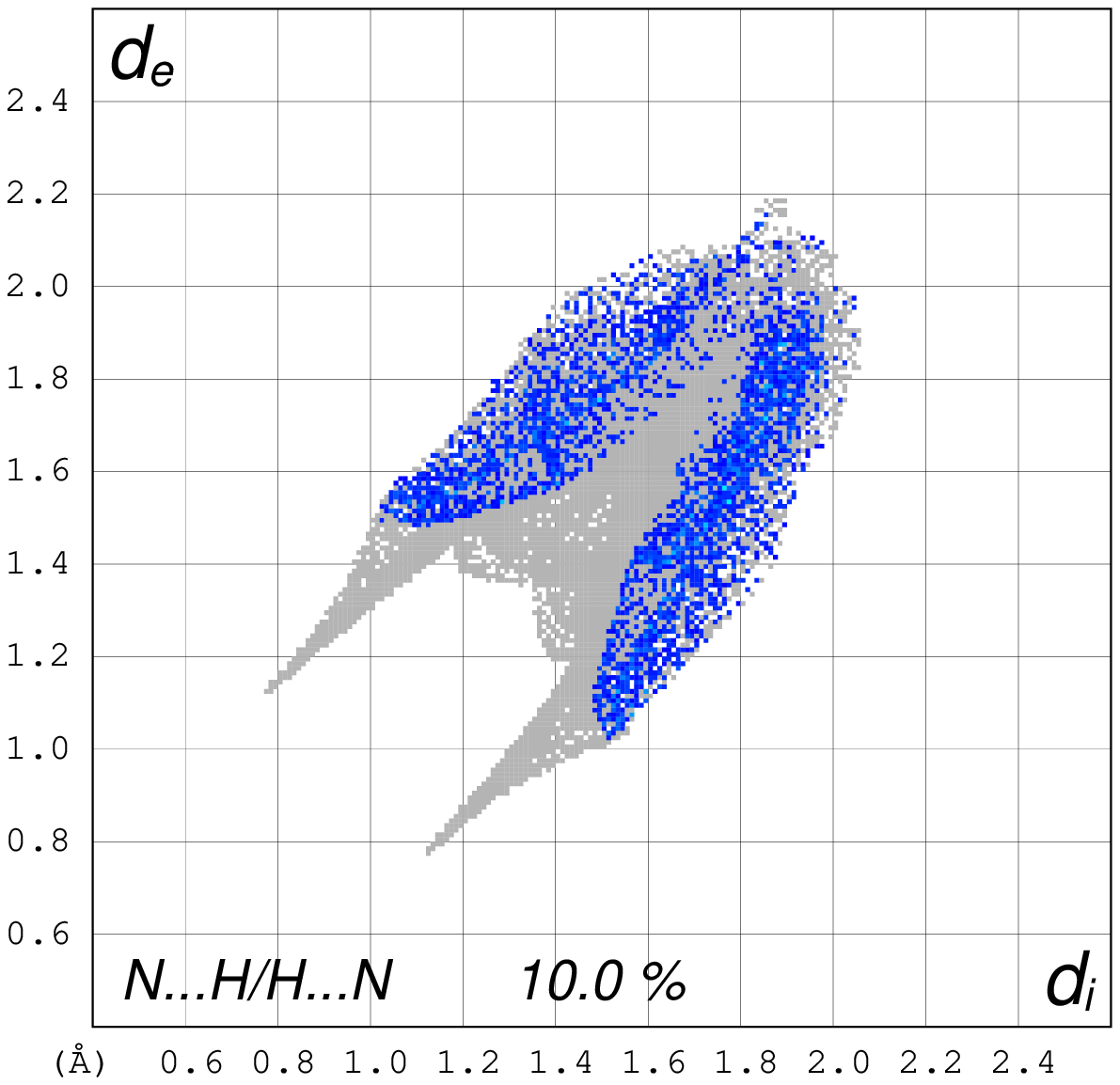}}} \hspace{0.1in}
{\subfigure[]{\includegraphics[height = 2.5in,width=2.5in]{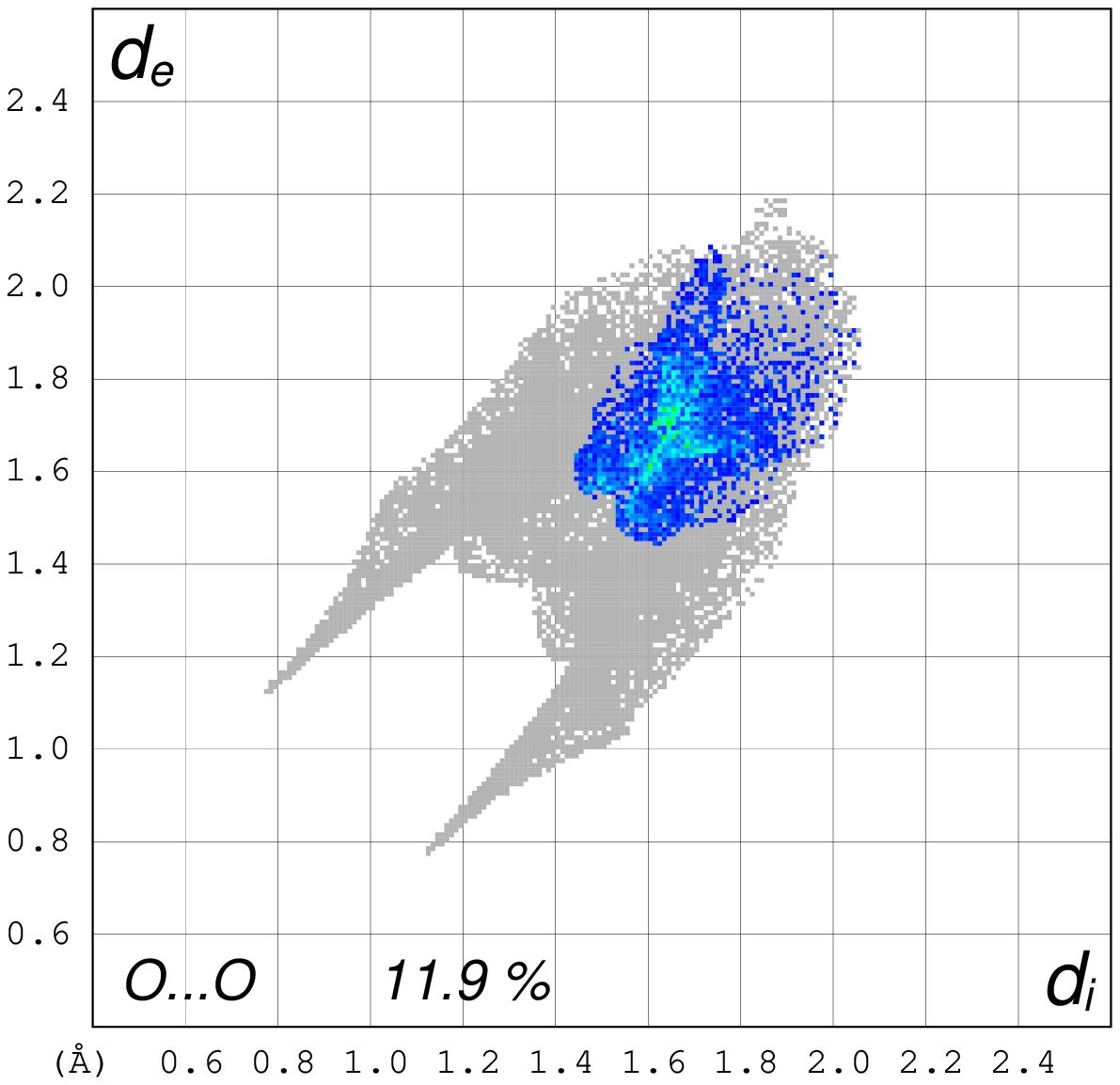}}}
{\subfigure[]{\includegraphics[height = 2.5in,width=2.5in]{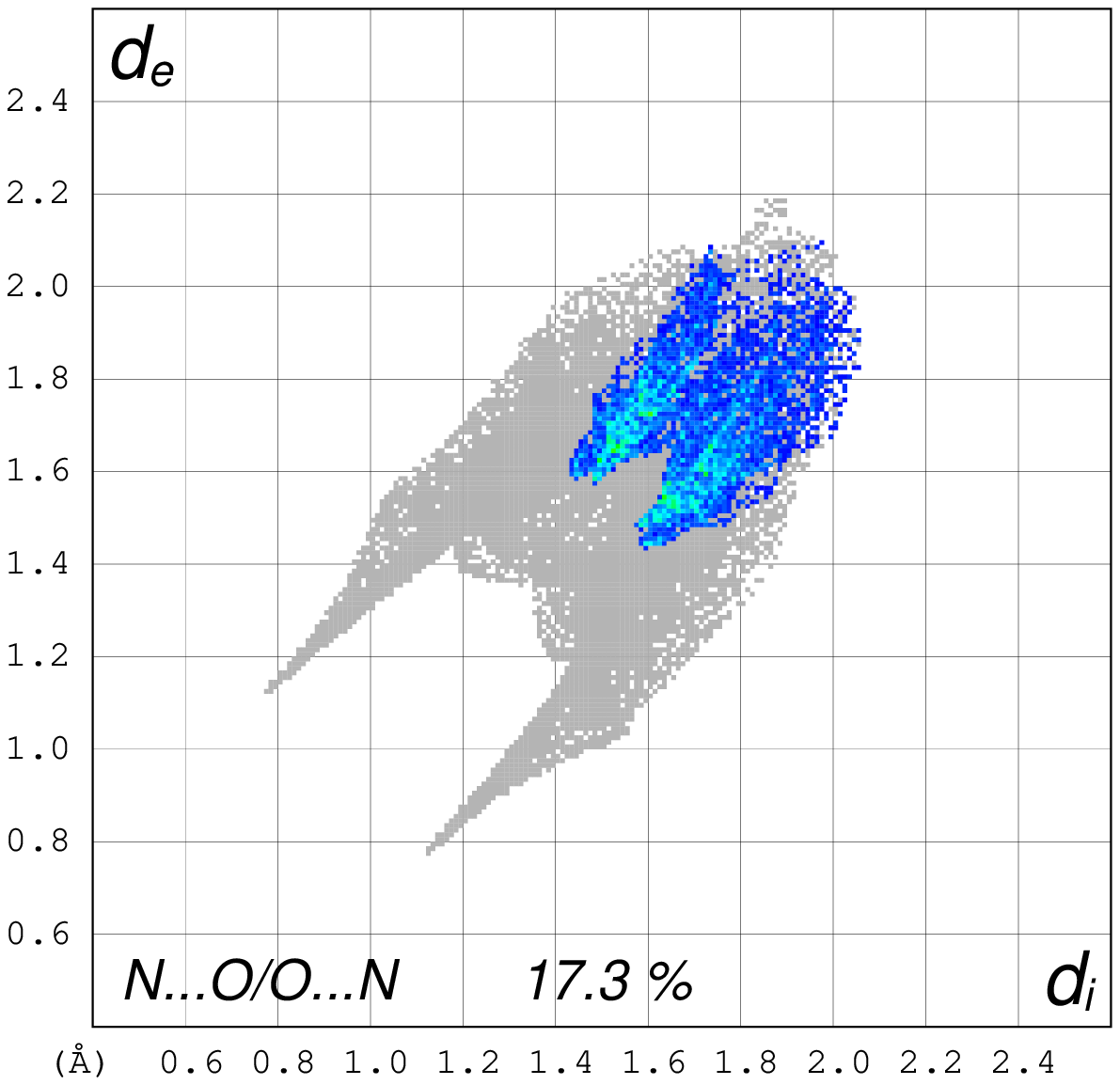}}}
\caption{(Color online) Calculated 2D finger print maps of Hirshfeld surfaces given in figure \ref{fig:HS} for ADN at ambient pressure.}
\label{fig:2D}
\end{figure*}

\begin{figure*}[h]
\centering
\includegraphics[height = 3.5in,width=5.5in]{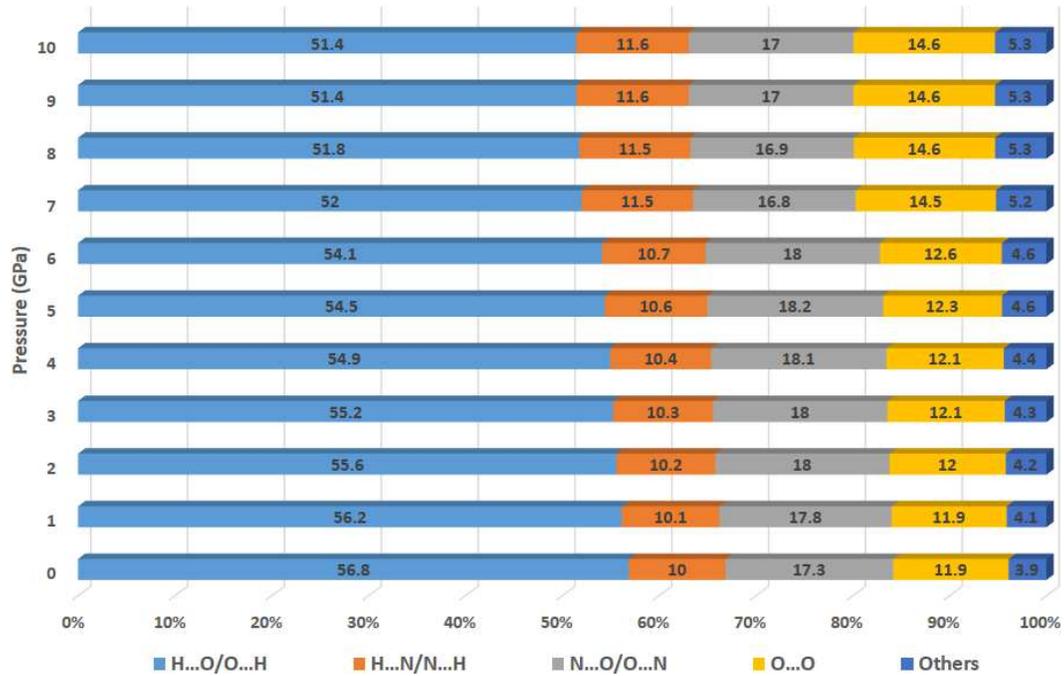}
\caption{(Color online) Percentage contributions to the Hirshfeld surface area for the various close intermolecular contacts for ADN as a function of pressure.}
\label{fig:HS-P}
\end{figure*}

\begin{figure*}[h]
\centering
{\subfigure[]{\includegraphics[height = 3.0in,width=4.0in]{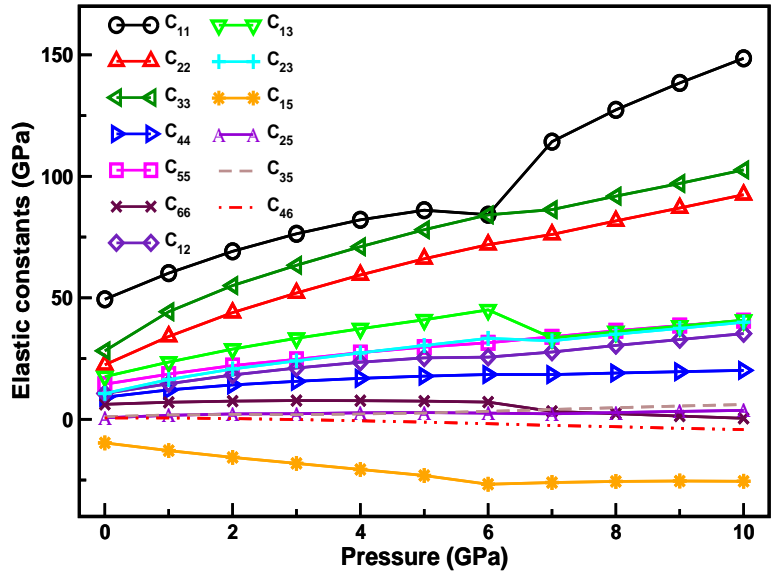}}}
{\subfigure[]{\includegraphics[height = 4.5in,width=3.5in]{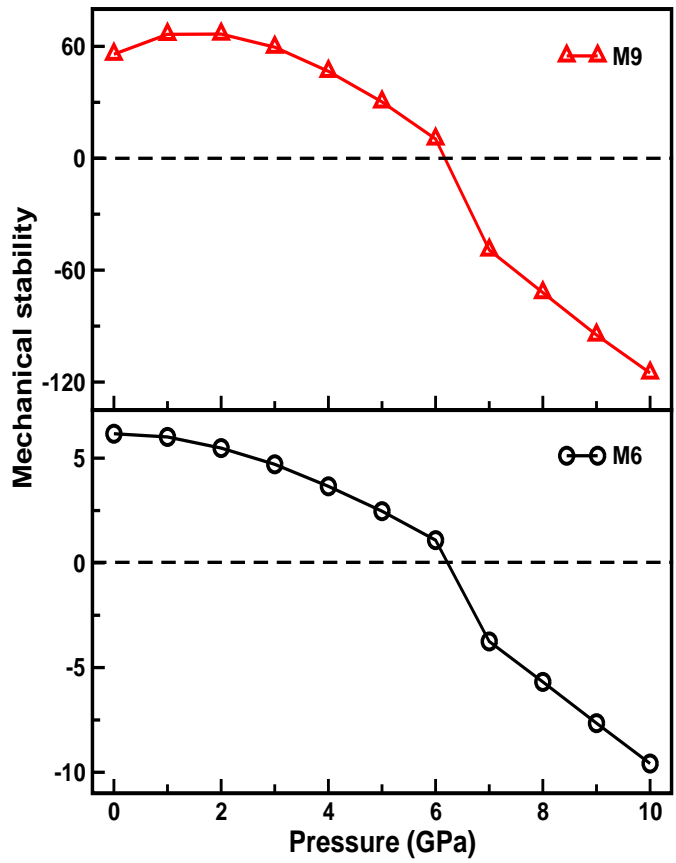}}}
\caption{(Color online) Calculated (a) elastic constants and (b) mechanical stability criteria of ADN as a function of pressure.}
\label{cij}
\end{figure*}

\begin{figure*}[h]
\centering
\includegraphics[height = 3.0in,width=4.5in]{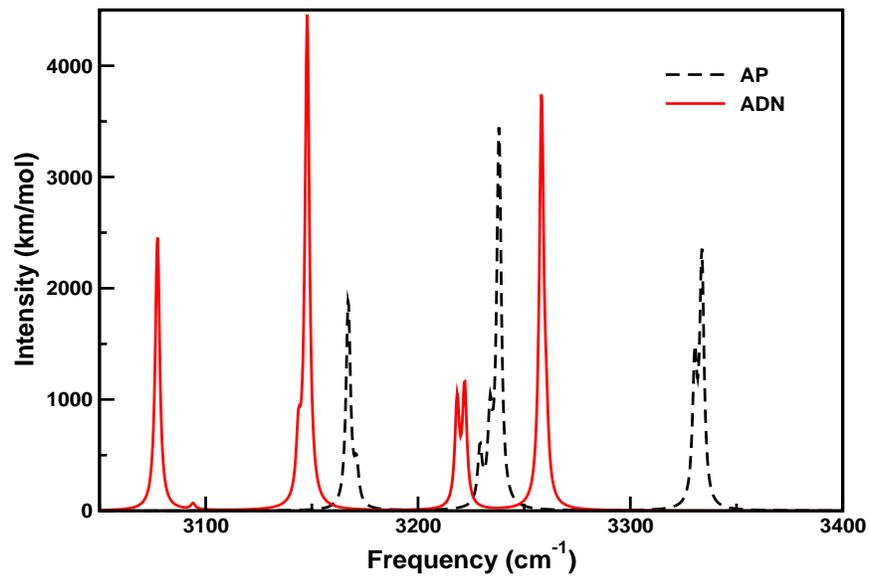}
\caption{(Color online) Calculated IR spectra of ADN is compared with AP in the mid-IR region at ambient pressure.}
\label{IR-0}
\end{figure*}

\begin{figure*}[h]
\centering
{\subfigure[]{\includegraphics[height = 2.6in, width=1.8in]{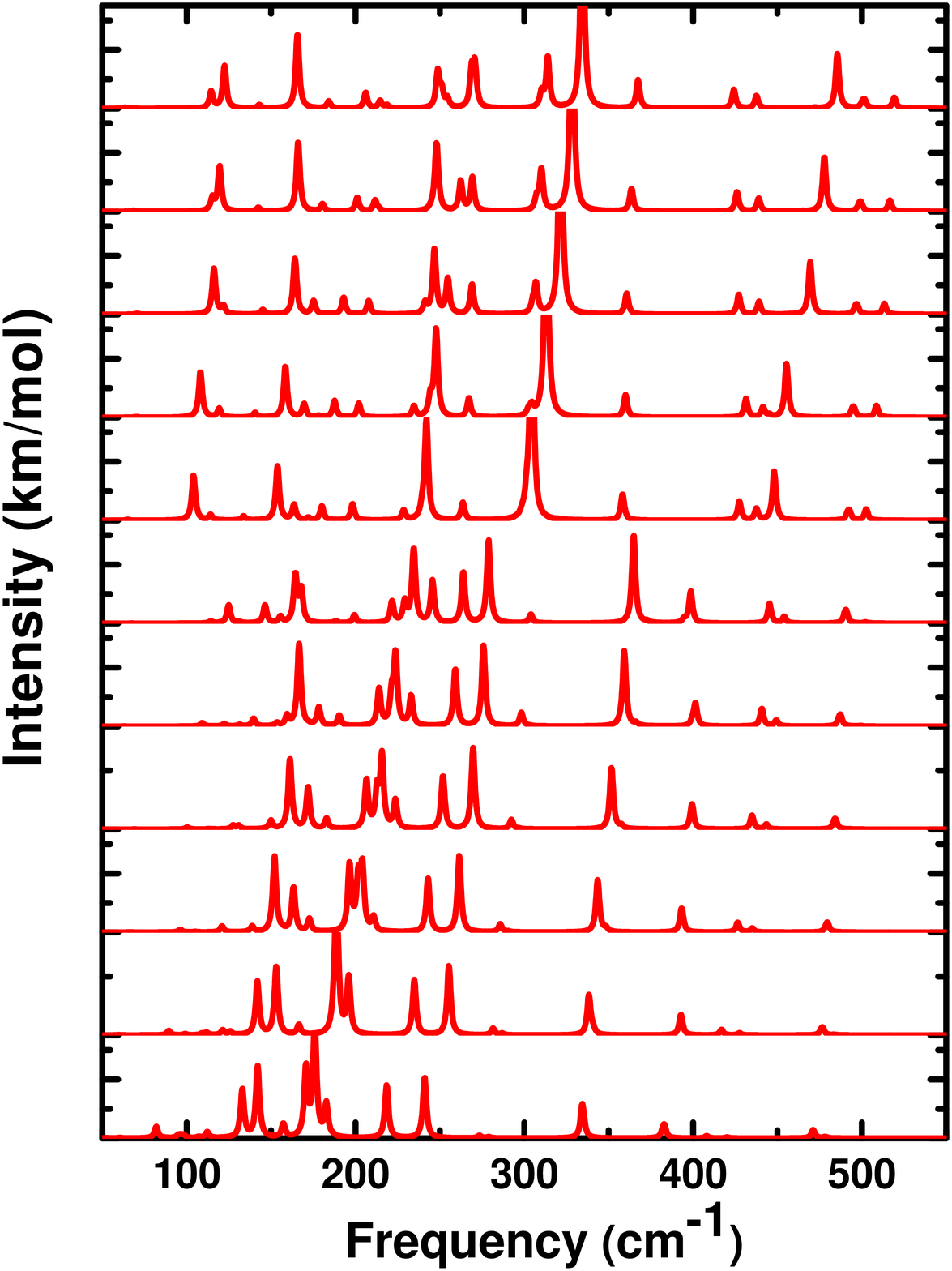}}}
{\subfigure[]{\includegraphics[height = 2.6in, width=1.8in]{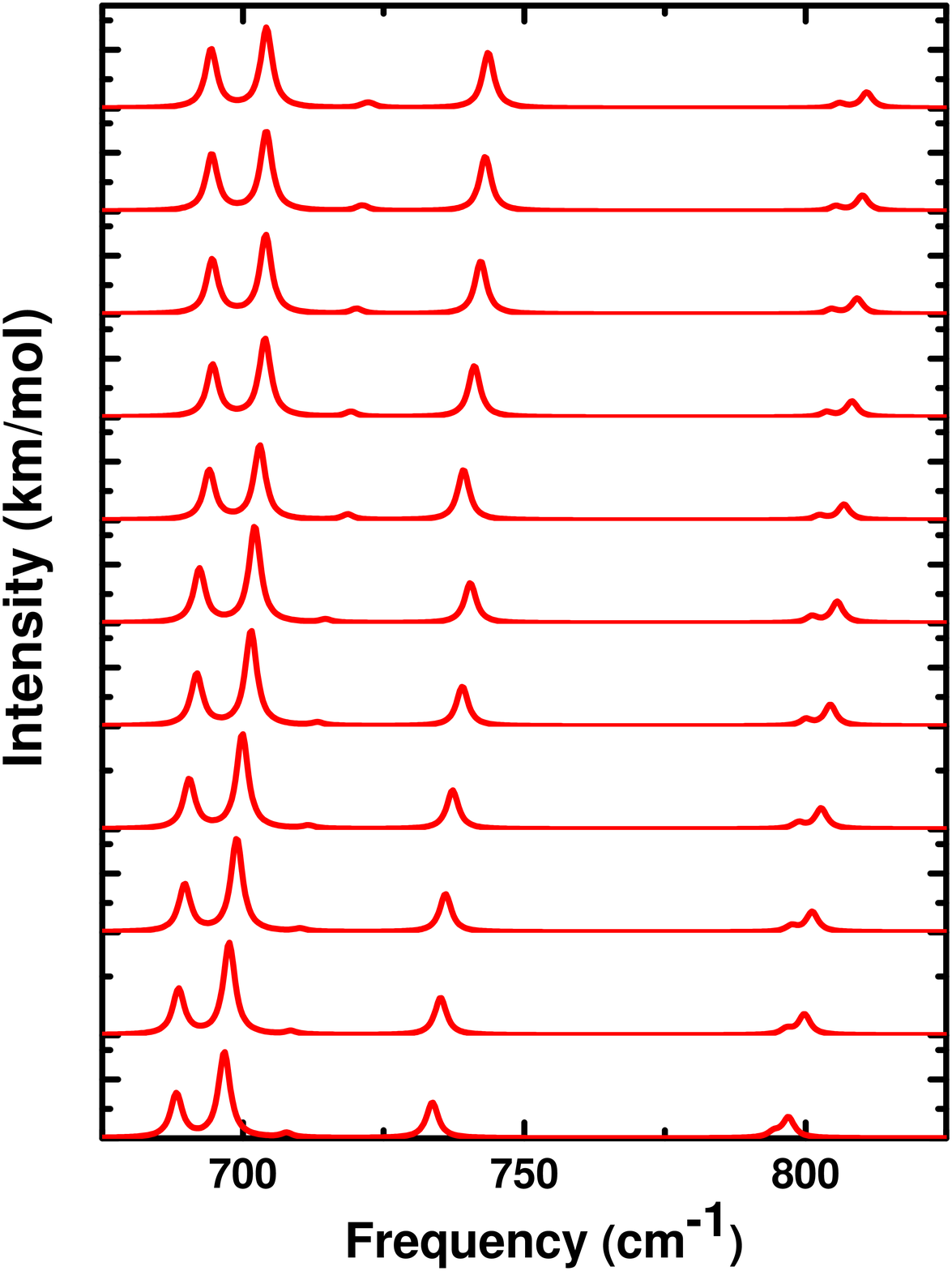}}}
{\subfigure[]{\includegraphics[height = 2.6in, width=1.8in]{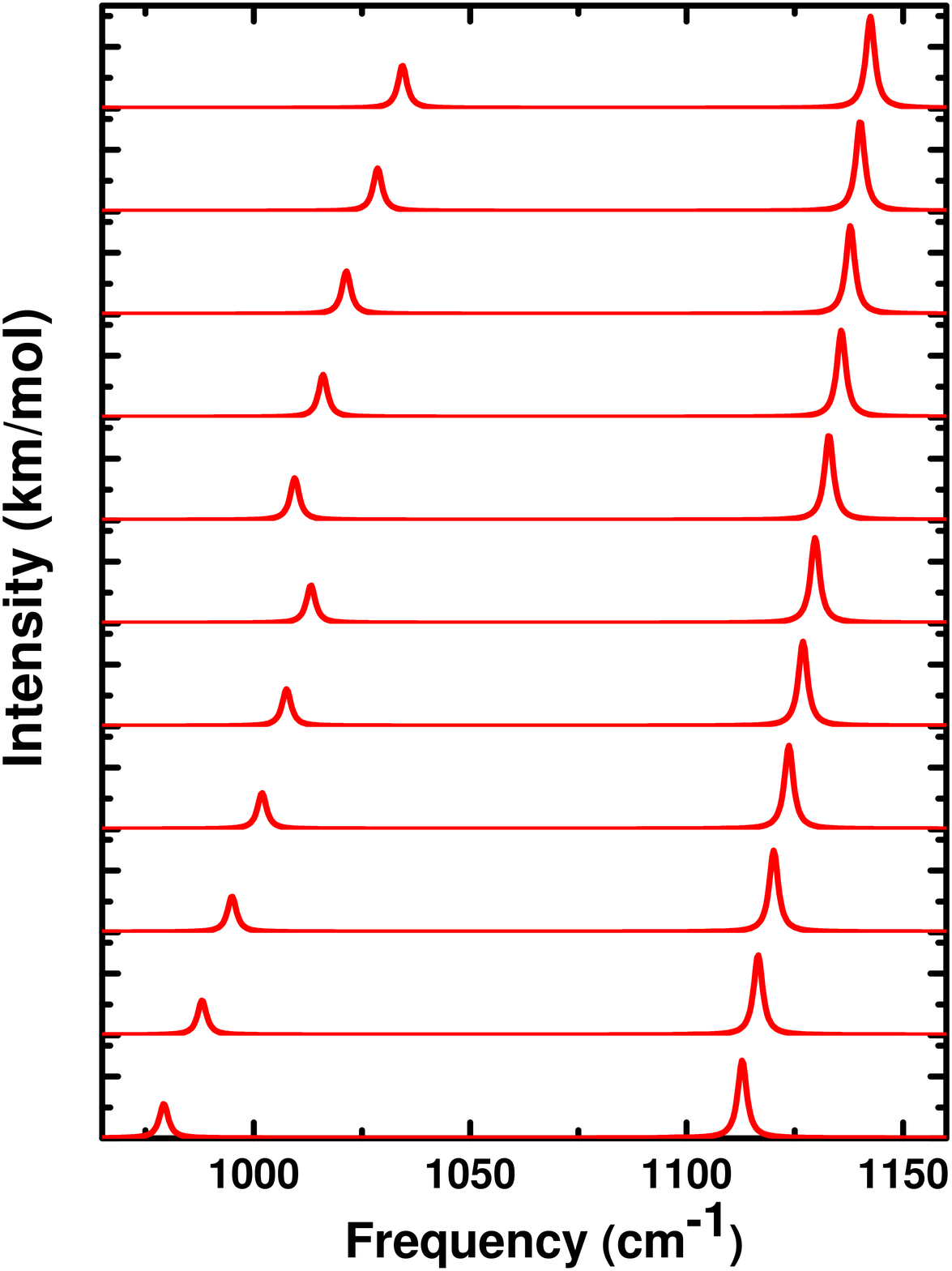}}}
{\subfigure[]{\includegraphics[height = 2.6in, width=1.8in]{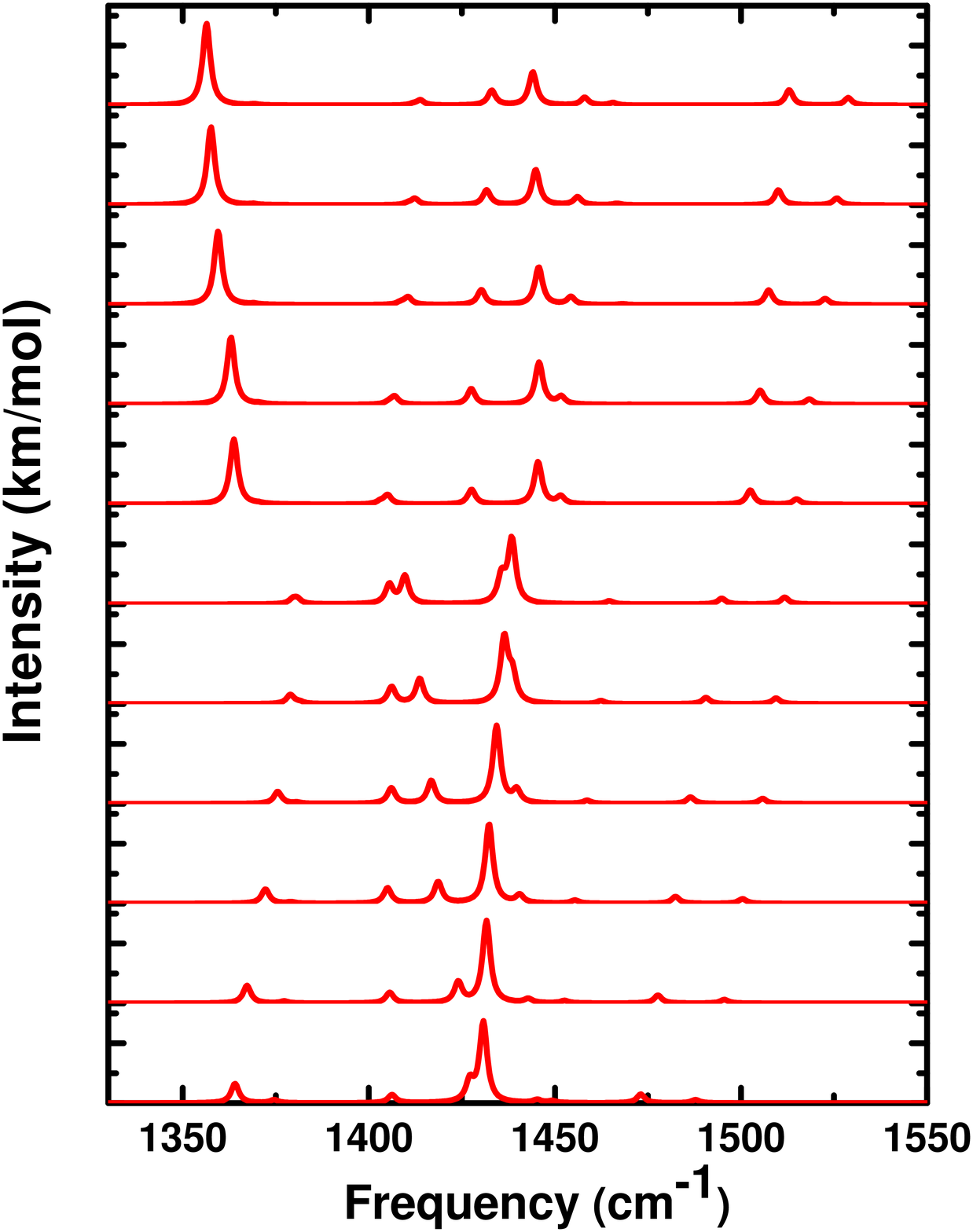}}}
{\subfigure[]{\includegraphics[height = 2.6in, width=1.8in]{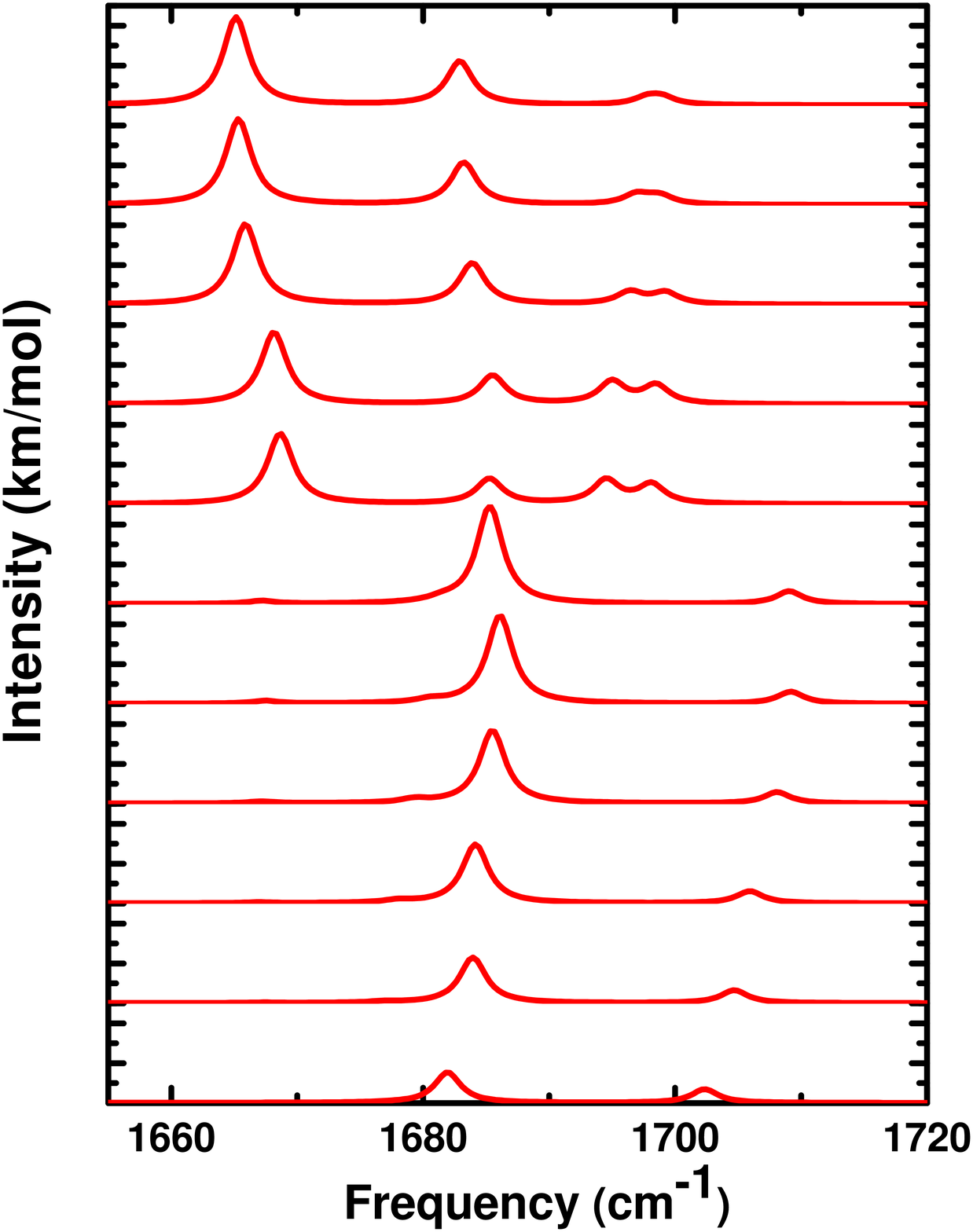}}}
{\subfigure[]{\includegraphics[height = 2.6in, width=1.8in]{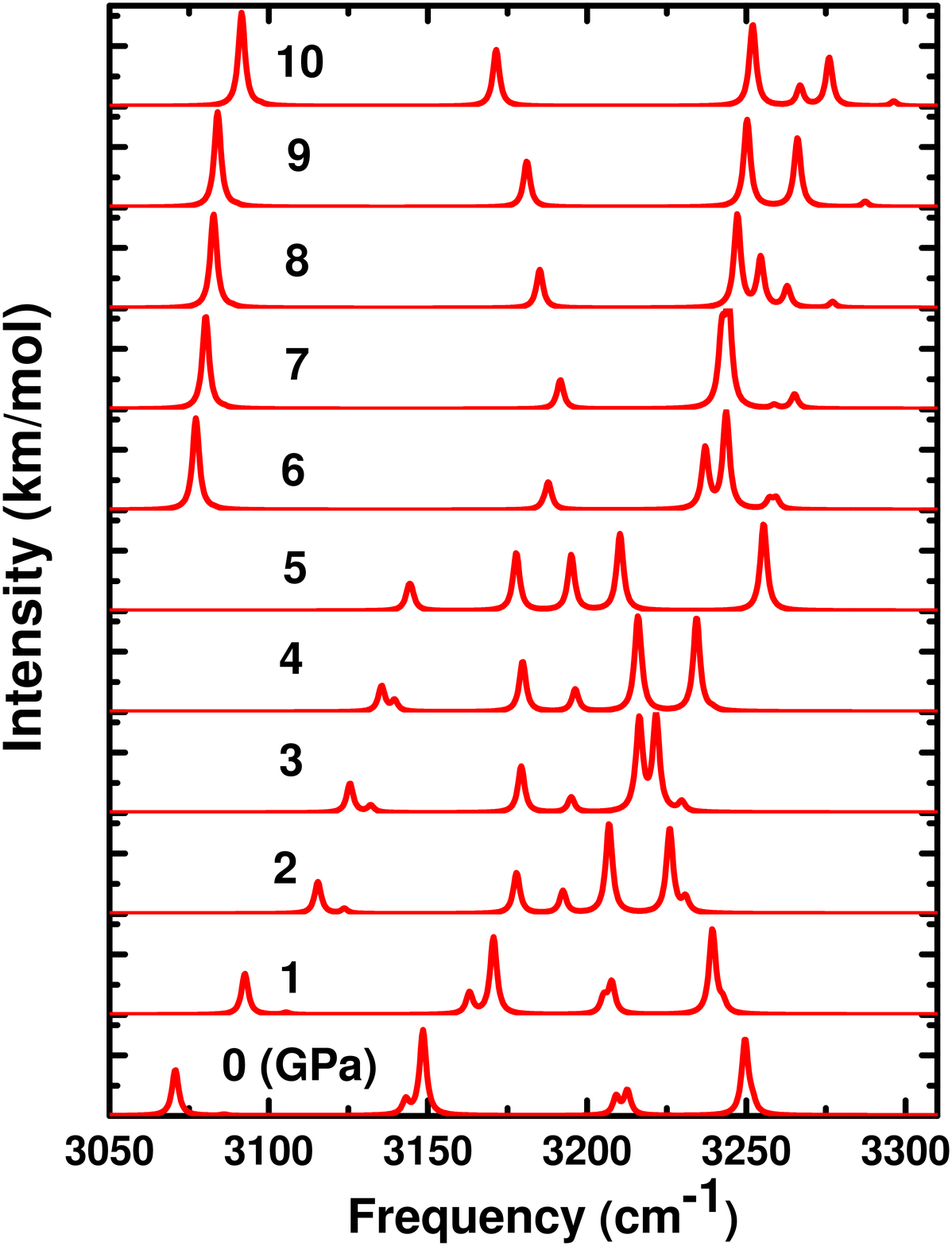}}}
\caption{(Color online) Calculated IR spectra (a) lattice modes (b) torsional and bending modes of NH$_4$ and N(NO$_2$)$_2$ ions (c) N(NO$_2$)$_2$ asymmetric modes (d) N-H wagging, rocking and scissoring modes (e) N-H bending modes (f) N-H symmetric and asymmetric stretching modes of ADN as a function of pressure.}
\label{IR2}
\end{figure*}

\clearpage
\begin{table*}[h]
\caption{Calculated equilibrium volume (V$_0$, in \AA$^3$), bulk modulus (B$_0$, in GPa) and its pressure derivative (B$_0'$) of ADN using standard PBE-GGA and various dispersion corrected (DFT-D) and non-local correction (vdW-DF) methods.}
\begin{ruledtabular}
\begin{tabular}{cccc}
 Method  &   V$_0$      &               B$_0$                                 &    B$_0'$   \\ \hline
  PBE    &   509.48     &               8.54                                  &    8.91      \\
  D2     &   450.03     &              18.48                                  &    8.20      \\
  TS     &   468.09     &              17.70                                  &    5.70      \\
 TS-SCS  &   472.83     &              16.20                                  &    6.14      \\
 vdW-DF  &   450.15     &               24.13                                 &    5.67      \\
 Others  &              &              20.65$^a$, 22.29$^{a*}$                &    4.03$^a$, 4.75$^{a*}$  \\
 Expt.   & 450.0$^b$    &              16.4$^c$                               &    6.5$^c$   \\         		
\end{tabular}
\end{ruledtabular}
\\
$^a$ fitting P-V data in the pressure range 0-10 GPa, Ref.\cite{Sorescu2}, \\
$^{a*}$ fitting P-V data in the pressure range 0-300 GPa, Ref.\cite{Sorescu2}, \\
$^b$Ref.\cite{Gilardi}, \\
$^c$Ref.\cite{Davidson3} \\
\label{tab:table1}
\end{table*}

\begin{table*}[h]
\caption{Calculated elastic constants (C$_{ij}$, in GPa) of ADN using DFT-D2 method.}
\begin{ruledtabular}
\begin{tabular}{ccccccccccccc}
C$_{11}$ & C$_{22}$ &  C$_{33}$ & C$_{44}$  & C$_{55}$  &  C$_{66}$  &  C$_{12}$ &  C$_{13}$ & C$_{23}$ &  C$_{15}$ & C$_{25}$ & C$_{35}$ & C$_{46}$                          \\ \hline
 49.4    &   22.4   &   28.2    &   9.1     &   14.4    &   6.2      &  10.6     &  17.6     &  10.7    &   -9.7    &   0.6  &  1.0 & 0.6 \\
\end{tabular}
\end{ruledtabular}
\label{tab:table2}
\end{table*}

\begin{table*}[h]
\caption{Calculated heat of formation (HOF, in kJ/mol), density ($\rho$), detonation velocity (D$_{CJ}$, in km/s) and pressure (P$_{CJ}$, in GPa) of AN, AP and ADN.}
\begin{ruledtabular}
\begin{tabular}{cccccccccc}
 Compound  &   Method  &            HOF                   &  $\rho$             &  D$_{CJ}$   &   P$_{CJ}$    \\ \hline
   AN      &  Present  &           -336                   &  1.734              &   7.28      &  18.71        \\
           &  Others   &    -326$^a$, -354.6$^b$          &  1.72$^b$           &             &               \\
   AP      &  Present  &            -236                  &  1.946              &   6.50      &  17.64        \\
           &  Others   & -295$^a$, -283.1$^b$, -298$^c$   & 1.95$^b$, 1.9$^c$   &             &               \\
   ADN     &  Present  &            -116                  &   1.831             &   8.09      &  25.54        \\
           &  Others   &    -149.7$^a$, -125.3$^b$,       & 1.81$^b$, 1.8$^c$   &   8.074$^b$ &  23.72$^b$    \\
           &           &    -151$^c$, -122.7$^d$          &                     &         &                   \\
\end{tabular}
\end{ruledtabular}
\\ $^a$Ref.\cite{Yang},
$^b$Ref.\cite{Kraue}
$^c$Ref.\cite{Nair},
$^d$Ref.\cite{Zeng}
\label{tab:table3}
\end{table*}
}
\end{document}